\documentstyle [aps,prd]{revtex}
\begin{document}
\input epsf
\def\la{\mathrel{\mathpalette\fun <}}
\def\ga{\mathrel{\mathpalette\fun >}}
\def\fun#1#2{\lower3.6pt\vbox{\baselineskip0pt\lineskip.9pt
  \ialign{$\mathsurround=0pt#1\hfil##\hfil$\crcr#2\crcr\sim\crcr}}}
\def\eg{{\it e.g.}}
\def\ie{{\it i.e.}}
\def\etal{{\it et al.~}}
\newcommand{\be}{\begin{equation}}
\newcommand{\ee}{\end{equation}}
\newcommand{\snr}{{\rm N_{S\!N}}\!(z)}
\newcommand{\msnr}{{\rm N_{S\!N}}\!(z,m)}
\newcommand{\snrx}{{\rm N_{S\!N}}\!(x)}
\newcommand{\lsnu}{\mbox{${\cal L}^{\rm S}_\nu$}}
\newcommand{\lsnuv}{\mbox{${\cal L}^{\rm S}_\nu\!(\epsilon^{\prime})$}}
\newcommand{\heat}{{\rm T}_\nu}
\newcommand{\yc}{\mbox{y$_{\rm cut}$}}
\newcommand{\mout}{\langle{\rm M_Z}\rangle}
\newcommand{\nbmout}{{\rm M_Z}}
\newcommand{\cm}{{\rm cm}}
\newcommand{\scminv}{1/{\rm cm}^2/{\rm s}}
\newcommand{\mevscminv}{1/{\rm cm}^2/{\rm s}/{\rm MeV}}
\newcommand{\aenu}{\langle{\rm E}_\nu\rangle}
\newcommand{\aheat}{\langle\heat\rangle}
\newcommand{\mev}{{\rm MeV}}
\newcommand{\antinue}{\bar{\nu}_{\rm e}}	
\newcommand{\etox}{\nu_{\rm e}\leftrightarrow\nu_{\rm x}}
\newcommand{\mutotau}{\nu_\mu\leftrightarrow\nu_\tau}
\newcommand{\ebtoxb}{\antinue\leftrightarrow\bar{\nu}_{\rm x}}
\def\snii{SN\thinspace{$\scriptstyle{\rm II}$}~}
\def\hi{H\thinspace{$\scriptstyle{\rm I}$}~}
\def\deriv#1#2{\frac{d #1}{d #2}}
\def\msun{{\rm M}_\odot}
\title{The Supernova Relic Neutrino Background}
\author{M. Kaplinghat$^{1,3}$ G. Steigman$^{1,2}$ and T. P. Walker$^{1,2}$}
\address{{\it 
$^1$Department of Physics, The Ohio State University, Columbus, OH 
43210\\ 
$^2$Department of Astronomy, The Ohio State University, Columbus, OH 
43210\\
$^3$Department of Astronomy \& Astrophysics, University of Chicago, 
Chicago, IL 60637}}
\maketitle
\begin{abstract}
An upper bound to the supernova relic neutrino background from all past  
Type II supernovae is obtained using observations of the Universal metal 
enrichment history.  
 We show that an unambiguous detection of these relic neutrinos by 
the Super-Kamiokande detector is unlikely. We also analyze the event rate
in the Sudbury Neutrino Observatory (where coincident neutrons from 
$\antinue D \rightarrow nne^+$ might enhance background rejection), and 
arrive at the same conclusion.
If the relic neutrino flux should be observed to exceed our upper bound
 and if the observations of the metal enrichment history 
(for $z<1$) are not in considerable error, 
then either the Type II supernova rate does not track the metal 
enrichment history or some mechanism may be responsible for 
transforming ${\bar {\nu}}_{\mu,\tau}\rightarrow \antinue$.  
\end{abstract}
\section{Introduction}

A Type II supernova (\snii) -- the explosion triggered by the gravitational 
collapse of a single massive star --  emits 99\% of its energy in neutrinos.  
The relic $\antinue$ background created by all past \snii is potentially
detectable in present and/or future large underground neutrino detectors. 
One of the goals of the current SuperK and SNO detectors is to detect this 
SRN background~\cite{totsuka,sno}. The predicted \snii relic $\antinue$ 
(SRN) flux depends crucially on the \snii rate as a function of redshift 
and the epoch of maximum SN rate (throughout the SN rate refers to the 
comoving densities of the \snii rate) is important in determining the 
detectability of the SRN background.  For example, if the SN rate should 
peak at redshifts of order 2 to 3, the majority of the SRNs would be 
redshifted to energies below typical detector threshold energies ($\sim$~5 
MeV).  Since the same objects which are responsible for creating the SRN 
background are also responsible for the bulk of the heavy element production, 
knowledge of the \snii metal and neutrino production, in concert with the 
observationally inferred metal enrichment history of the Universe, is the
straightest path (\ie, least model dependent) to predicting the flux and 
spectrum of relic $\antinue$ from all past \snii \footnote{Type I supernovae 
are not expected to contribute appreciably to the relic neutrino background.}.  
It is our goal here to follow this 
path in providing a generous upper bound to the expected SRN background.  
Furthermore, we account for the characteristics of the SuperK and SNO 
detectors, and use our calculated (upper bound to the) SRN background 
to predict (upper bounds to) the event rates at these detectors.  These 
event rates are compared to expected backgrounds and to current limits. 
The current upper limit~\cite{zhang} (from the Kamiokande II detector) 
on the flux of supernova relic $\antinue$ in the energy interval from 
19 to 35 MeV is 226~${\rm cm}^{-2}{\rm sec}^{-1}$.  SNO is just beginning 
operation and has not decided upon a neutron detection strategy which is 
vital to the detection of $\antinue$. 

In the last few years progress has been made in constraining the recent 
star formation history of the Universe.  In particular, a variety of 
observational evidence seem consistent with a comoving star formation 
rate (SFR) density which was much higher at redshifts $z \sim 1$ than 
at the present epoch.  The history of star formation beyond $z \sim 1$ 
is less certain and it is not yet clear if the Universal SFR declined 
rapidly or evolved only mildly at higher redshifts.  Support for this 
scenario comes from Pei \& Fall~\cite{pei} who have used chemical evolution 
models to explore the SFR and metal enrichment history inferred from 
observations of damped Ly$\alpha$ systems.  They find that the observed 
\hi column densities may not represent the true column densities 
because of significant corrections due to dust.  Since the Ly$\alpha$ 
systems are identified from the spectra of quasars and since the Ly$\alpha$ 
systems may contain dust, the implication is that some of the quasars may 
be invisible.  This, in turn, suggests that some Ly$\alpha$ systems may 
go undetected.  When Pei \& Fall correct for the effects of dust obscuration, 
they find evidence for rapid star formation at low redshifts ($z \sim 1$).  
This is to be compared to the predicted peak star formation rate epoch at 
redshifts of order 3 -- 4 when obscuration is not taken into account.  
In particular, for their model with infall Pei \& Fall find that the 
observational data is consistent with a SFR which increases until $z \sim 1$ 
and then decreases with further increases in redshift.  Independent 
observations by Madau \etal \cite{madau} of the metallicity enrichment rate 
(MER) are in excellent agreement with the Pei \& Fall results.  The direct 
quantitative support for Pei \& Fall's model comes from the Canada-France 
redshift survey of faint galaxies~\cite{lilly} which found that the 
comoving UV luminosity density of the Universe shows a sharp decline 
from $z\sim 1$ to the present. 

Here, using the model of Pei and Fall, we parameterize the metal enrichment
history (MER) observed by Madau \etal \cite{madau} to predict the
\snii relic $\antinue$ flux at Earth.  In the past, similar calculations 
of this relic $\antinue$ flux have been done by Totani \& Sato \cite{tot-sato}, 
Totani, Sato \& Yoshii \cite{totani}, Malaney \cite{malaney} and Hartmann \& 
Woosley \cite{hartmann}.  In contrast to almost all of the above analyses, we 
strive to minimize any model dependences by directly relating the supernova
rate and its evolution to observations of the metal enrichment history to 
obtain supernova rate and, thereby, the SRN flux.  In our calculations we 
make always make ``conservative" choices of any uncertain parameters so as 
to obtain a robust {\it upper} bound to the SRN flux.  From this upper bound 
we will conclude that it is unlikely SuperK will detect these relic neutrinos 
(because the signal will be buried under a large background event rate) and 
that the event rate in SNO should be vanishingly small. 
In \S\ref{flux}, we outline the formalism for calculating 
the flux of SRN at Earth.  
In \S\ref{event-rate}, we calculate conservative 
(\ie, generous) {\it upper} bounds to the relic $\antinue$ event rates at 
SuperK and SNO. 
In \S\ref{discuss}, we review previous estimates of the SRN flux in 
comparison with ours and we examine neutrino oscillations as a possible 
mechanism for increasing the SN relic $\antinue$ flux.  

\section{The Supernovae Relic Neutrino Spectrum }
\label{flux}

The spectrum of neutrinos at Earth due to all past supernovae depends 
on the differential (per unit energy interval) neutrino flux from each 
SN, on the redshift distribution of the SN rate, and on an assumed 
Friedmann-Robertson-Walker cosmology which may be parameterized by the 
Hubble parameter $H_0$ and the matter density parameter $\Omega_0$.  For
simplicity we ignore a possible cosmological constant at this point and 
discuss its effect later.  If the supernova rate per unit comoving volume 
at redshift $z$ is $\snr$ and the neutrino energy distribution at the source 
(at energy $\epsilon$) is $\lsnu\!(\epsilon)$, then the differential flux 
of relic neutrinos at Earth is given by 
\be
\label{current}
j_{\nu}(\epsilon)=
\frac{c}{H_0}\int_0^{\infty}
dz\;\frac{\snr\,\langle\lsnuv\rangle}{(1+z)\,\sqrt{1+\Omega_{0}z}}\,,
\ee
where \(\epsilon^{\prime}=(1+z)\epsilon\) and the neutrinos are
assumed to be massless. The angled brackets indicate that the
dependence of the neutrino flux on supernova progenitors with
different masses should be averaged over the initial mass function
(IMF). In practice we will choose values for these average quantities
so as to {\it maximize} the SRN background. 

The spectrum of the neutrinos from a supernova is parameterized as 
a Fermi-Dirac distribution with zero chemical potential, normalized to 
the total energy in a particular neutrino species (${\rm E}_\nu$) emitted 
by the supernova,
\ie, $\int \lsnu\!(\epsilon)\epsilon d\epsilon={\rm E}_\nu$.  
Then, for each neutrino species $\nu$,
\be\lsnu\!(\epsilon)={\rm E}_\nu\times\frac{120}{7\pi^4}\times
\frac{\epsilon^2}{\heat^4}\times\left[\exp\left(\frac{\textstyle \epsilon}
{\textstyle \heat}\right)+ 1\right]^{-1}.\ee 
The neutrino luminosity is thus characterized by ${\rm E}_\nu$ and 
$\heat$ which, in turn, depend on the SN progenitor mass.  However, 
the problem of obtaining the IMF-averaged neutrino flux simplifies 
because $\heat$ does not vary rapidly as the SN progenitor mass is 
changed~\cite{woosley}.  Adopting a flat $\Omega_0=1$ cosmology, 
and setting $x \equiv 1+z$ , we can then write eq.~\ref{current} 
to a good approximation as
\be
j_{\nu}(\epsilon)={\cal A}\,\frac{\aenu}{\aheat^4}\, 
\epsilon^2\,\int_{1}^{\infty}dx\;
{\snrx}\,\frac{\sqrt{x}}{\exp(\epsilon x/\aheat)+1},
\label{djdy}
\ee 
where ${\cal A}=(120/7\pi^4)\,cH_0^{-1}= 1056\,h_{50}^{-1}$ Mpc, 
with $H_0=50\,h_{50}$~km/s/Mpc. 
The results of Woosley \etal\cite{woosley} for a $25~{\rm M}_\odot$ 
supernova progenitor are used to fix $\aenu=11\times 10^{52}~{\rm ergs}$ 
and $\aheat=5.3~\mev$.  The values of $\aenu$ and $\aheat$ characterize 
the detectable $\antinue$ supernova neutrino spectrum . 
For comparison, recall that the data from SN 1987A gave ${\rm E}_\nu=8
\times 10^{52}~{\rm ergs}$ and $\heat=4.8~\mev$~\cite{sn1987a}.  Another 
issue concerns the role of the IMF and selection of the $\aenu$ and $\aheat$ 
values as averages.  To obtain an upper bound to the detection rate, 
we use values of $\aenu$ and $\aheat$ which provide upper bounds to 
any reasonable average. This is possible because the flux integrated 
over the {\em observable} energy window (and hence the event rate) is an 
increasing function of both $\aenu$ and $\aheat$.  The value of $\heat$ 
is particularly insensitive to the progenitor mass because $\heat$ 
derives its value from the temperature of the neutrinosphere formed 
during the collapse and the thermodynamic properties of the neutrinosphere 
(as long as it is well-defined) do not vary much with mass~\cite{woosley}.  
In particular, $\heat\sim 4-5~\mev$ with 5.3~MeV being at the the upper 
end of the range. Thus, guided by models and SN 1987A, we parameterize 
the neutrino flux from supernovae so as to obtain a conservative upper 
bound to the SRN event rate.  

Under the assumption that the supernova rate tracks the metal enrichment
rate, the supernova rate used to calculate the relic neutrino flux can be 
written as
\be
\label{nsn}\snr=\frac{\dot{\rho}_{\rm Z}(z)}{\mout},
\ee
where $\mout$ is the average yield of ``metals" (${\rm Z}\geq 6$) 
per supernova and $\dot{\rho}_{\rm Z}$ is the metal 
enrichment rate per unit comoving volume.  
We have implicitly assumed that the metals come from \snii 
consistent with nucleosynthesis arguments~\cite{arnett} which show 
that the metal enrichment role of SN Ia is secondary to that of \snii. 
In any case, by neglecting SN Ia we overestimate the SRN flux (albeit, 
by only a factor of 2 at most). The other point to be noted here 
concerns our use of the metal enrichment history instead of the 
possibly more direct SFR to compute the SN rate. Both the SFR and 
the metal enrichment rate can be inferred from observations of the 
UV luminosity of star forming galaxies. But unlike the SFR which has 
a steep dependence on the adopted IMF, the metal enrichment rate is 
less sensitive to the IMF because the same (heavier) stars which are 
more UV luminous also eject more metals~\cite{madau}.  Thus, the SN 
rate is more closely tied to the metal enrichment history than to 
the SFR. 

To parameterize the evolution of $\dot{\rho}_{\rm Z}(z)$ from the present 
back to $z=1$, we use the results of Pei \& Fall.  In particular, we 
use the comoving metal production rate for their case with infall 
(Fig.1 of~\cite{fall}) which is in good quantitative agreement with 
SFR observations at $z<1$~\cite{lilly,connolly,gallego,madau}.  Since 
the neutrino flux from individual supernovae falls rapidly with 
increasing energy, and the lower energy neutrinos from high redshift 
supernovae are redshifted below the threshhold of detectability, our 
predictions are relatively insensitive to the high redshift ($z > 1$) 
behavior (as also noted by Hartmann \& Woosley~\cite{hartmann}).  
This is fortunate since it is difficult to quantify precisely 
the $z>1$ evolution.  For these reasons, we make the simplifying 
conservative assumption that the supernovae rate remains constant at 
higher redshifts: ${\rm N_{S\!N}}(z>1)={\rm N_{S\!N}}(z=1)$.  It should 
be noted that the $z<1$ evolution adopted here is likely quite robust 
in that independent studies reveal the same pattern of evolution 
(including, for example, that of the QSO luminosity density~\cite
{boyle}) and the different observational data are in good quantitative 
agreement.  Nevertheless, some changes to our adopted chemical enrichment 
history could be envisaged 
based on the arguments that the role of dust at high redshifts is still 
uncertain and, perhaps not all the star formation at higher redshifts 
has been observed~\cite{pascarelle,hughes,meurer,steidel}. However, the 
relative insensitivity of our upper bound to the high redshift behavior 
insulates it against such uncertainty. 

To determine the average amount of metals ejected per supernova, the 
results of the calculations of supernova nucleosynthesis by Woosley 
and Weaver \cite{ww} are employed. From their published tables of the 
elemental composition of the ejecta, it can be ascertained that the 
heavy element yield ranges from $\nbmout=1.1~\msun$ for a $15~\msun$ 
SN progenitor to $\nbmout=4.2~\msun$ for $25~\msun$. These results 
are for an initial metallicity equal to $0.1~{\rm Z}_\odot$, which we 
assume characterizes the metallicity at redshifts around unity \cite{smith}. 
In any case, the $\nbmout$ values for SN progenitors with initial 
metallicity equal to ${\rm Z}_\odot$ is greater by about 10-20$\%$ 
which, if used, would lead to a {\it decrease} in the predicted flux.  
Keeping in mind that the rate of events varies inversely as $\mout$,
we set $\mout$ equal to $1~\msun$ in the interest of obtaining an 
unambiguous upper bound.  In Figure~\ref{spectrum} we show the SRN 
spectrum that results from our adopted metal enrichment history and a 
conservative lower bound to the SN metallicity yields, $\mout=1~\msun$.

\section{Event rate at SuperK and SNO}\label{event-rate}

\begin{figure}[Fig1]
\centering
\leavevmode\epsfxsize=12cm \epsfysize=15cm\epsfbox{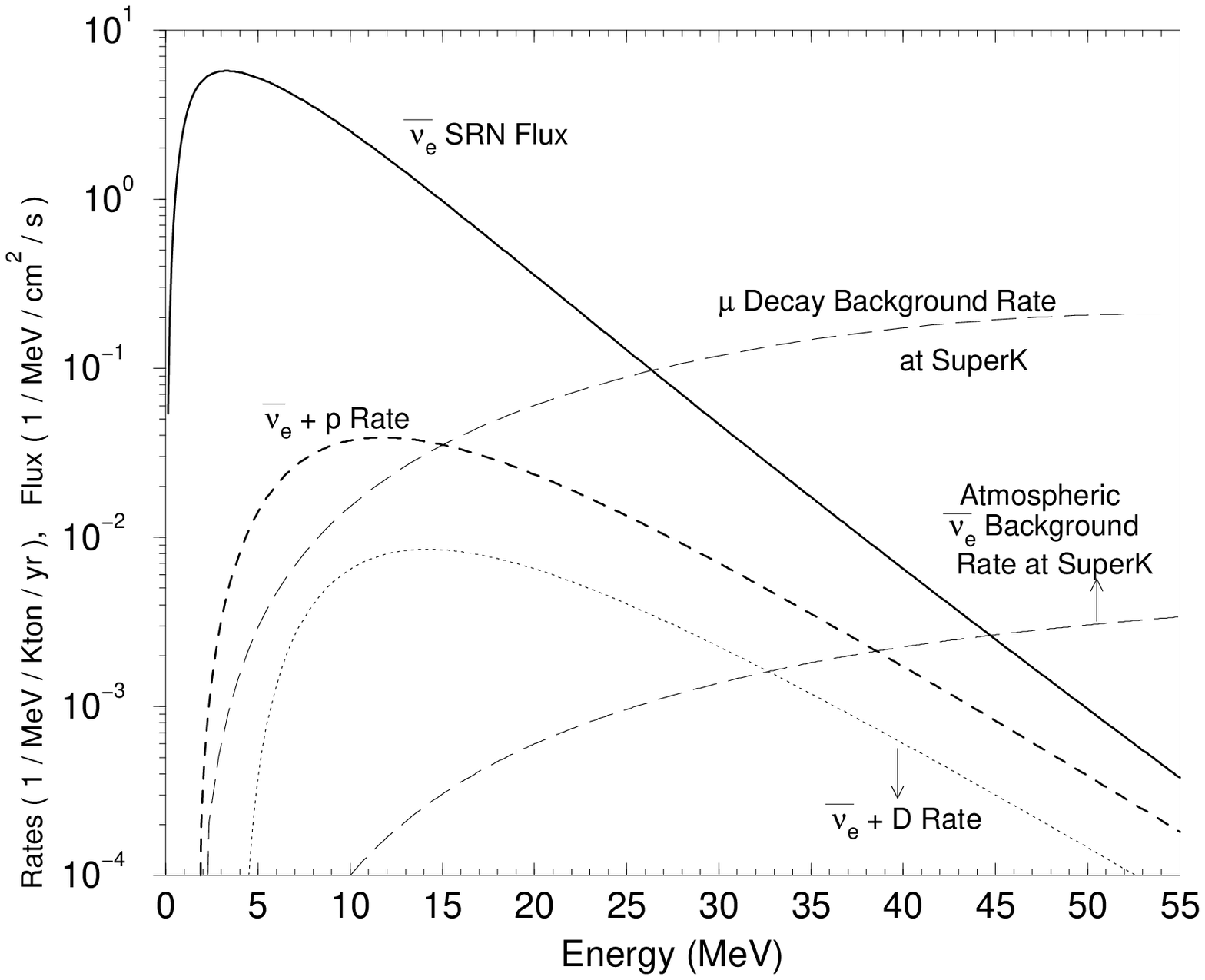}\\
\
\caption[Fig1]{\label{spectrum} The relevant event rates for the 
detection of supernova relic neutrinos (SRN), along with the predicted 
SRN spectral flux. The $\antinue p$ rate is for SRN detection
at SuperK, while the $\antinue D$ rate applies for the SNO detector. 
The abscissa, energy, refers to the $\antinue$ energy for all the 
cases {\it except} for the $\mu$ decay rate where it corresponds to 
the decay-produced electron's energy (${\rm E_e}$) plus 1.3 MeV 
(\ie, what the energy of a $\antinue$ would have to be, in order to 
produce a positron with energy ${\rm E_e}$ by $\antinue+p$ reaction.)}
\end{figure}

It is not possible to detect SN relic neutrinos at all energies.  For SuperK 
the observable energy window is likely to be from 19 to 35 MeV.  Below 10 
MeV, the $\antinue$ from reactors~\cite{lagage} and the Earth will completely
overwhelm the relic neutrinos. Above 10 MeV and below the observable energy 
window, the main source of background is due to the solar neutrinos, radiation 
from outside the fiducial volume and spallation-produced events due to the 
cosmic-ray muons in the detector~\cite{zhang}. Above 19 MeV the background is 
primarily due to atmospheric neutrinos~\cite{gaisser}. At energies greater 
than about 35 MeV, the rapidly (exponentially) falling SRN flux (peaked around 
3 MeV) becomes smaller than the atmospheric $\antinue$ flux, as can be 
verified from Figure \ref{spectrum}.  Therefore the observable flux is 
obtained by integrating the differential flux over the {\em neutrino} 
energy range from 20.3 to 36.3 MeV (since $\epsilon=E_e+1.3$ MeV 
where $E_e$ is the energy of the positron and 1.3 MeV is the 
neutron -- proton mass difference). 
We will also quote results in the more optimistic energy window of 
15 -- 35 MeV in the hope that with better background subtraction, SuperK 
will be able to probe these lower energy relic neutrinos.
Detection of the SRN background in the much smaller SNO detector may be 
possible using coincident neutrons from $\antinue D \rightarrow nne^+$.  
Because neutron detection at SNO is still in its infancy, we quote the 
total SNO event rate for positron energies above 10 MeV, corresponding 
to our SRN background.

To calculate the event rate at SuperK, the detector is assumed to be 
100$\%$ efficient in the observable energy window.  The dominant reaction 
is $\antinue p\rightarrow n e^+$ with a cross section ($\sigma_p(\epsilon)$) 
two orders of magnitude larger than that of the scattering reaction 
($\nu_{\rm e}e\rightarrow\nu_{\rm e}e$).  The differential event rate
in the interval $d\epsilon$ is then $N_p\sigma_p(\epsilon)j_\nu(\epsilon)
d\epsilon$ and the predicted event rate at the detector is: 
\be
{\rm R}={\cal A}\,{\rm N_p}\,\frac{\aenu}{\aheat^4}\,
\int_{1}^{\infty}dx\;\snrx \sqrt{x}\,
\int_{\epsilon_1}^{\epsilon_2}d\epsilon\;\frac{\epsilon^2\,\sigma_p(\epsilon)}
{\exp(x\epsilon/\aheat)+1}\;\;,
\label{rate}
\ee
where the $\epsilon_i$ delineate the energy window (for this case, 20.3 and 
36.3, respectively)
and ${\rm N_p}$ is the number of free protons in the detector. For SuperK, 
with a fiducial volume of 22.5 ktons, ${\rm N_p}=1.51\times10^{33}$.

Using the metal enrichment history to establish the supernova rate, 
the SN relic $\antinue$ event rate at SuperK can be written as
\be
{\rm R} = 0.066\,\left(\frac{\msun}{\mout}\right)\,
\left(\frac{\aenu}{10^{52}\,{\rm ergs}}\right)\,
\left(\frac{\aheat}{{\rm MeV}}\right)\;
\frac{\rm events}{\rm 22.5\;kton\!-\!year}
\ee
where we use $\sigma_p(\epsilon) = 9.52\times 10^{-44}\, E_e\, p_e~\cm^2$
\cite{vogel+beacom} 
with $E_e$ and $p_e$ (the energy and momentum of the positron) measured 
in ${\rm MeV}$.  We have set $h_{50}=1$ in the interest of obtaining an 
upper bound to the event rate.  Also, for the same reason the average 
metal yield per supernova is taken to be $1~\msun$, a lower bound to that 
obtained in the Woosley \& Weaver models. For completeness we show in 
Figure~\ref{spectrum} the differential rate of $\antinue p\rightarrow 
n e^+$ for our SRN background, with $\aenu=11\times 10^{52}~{\rm ergs}$ 
and $\aheat=5.3~{\rm MeV}$.

With our adopted SN parameters, the SRN event rate for a 22.5 kton-year 
exposure at SuperK is predicted to be 
\begin{eqnarray}
{\rm R} &<& 4~{\rm events}  
\qquad 19 \le {\rm E_e (MeV)} \le 35 \,, \nonumber \\
{\rm R} &<& 7~{\rm events} 
\qquad 15 \le {\rm E_e (MeV)} \le 35 \,.
\label{result}
\end{eqnarray}
Because the SRN spectrum falls rapidly with energy, the energy 
distribution of the events is strongly peaked at about 10 MeV 
(see Fig. 1; in 5 MeV bins from 10 MeV to 40 MeV, the percentages 
are 37:29:17:10:5:2). If the threshold could be lowered to 10 MeV, 
our upper bound to the event rate at SuperK would increase to 
about 10/year. 
In terms of the flux at the detector, the results are as follows: 
the upper bound to the SRN flux integrated over all energies 
is $54~{\rm {\mbox cm^{-2}sec^{-1}}}$
while in the relevant energy window from 19 to 35 MeV, the flux is 
$1.6~{\rm {\mbox cm^{-2}sec^{-1}}}$ 
(to be compared to the current \cite{zhang} upper bound 
of $226~{\rm {\mbox cm^{-2}sec^{-1}}}$). In the larger
energy window from 15 to 35 MeV, the observable flux is
$3.7~{\rm {\mbox cm^{-2}sec^{-1}}}$. The reason 
for the large difference between the total and observable flux 
is two-fold. One, the observable energy window only captures the 
falling tail-end of the SRN spectrum and two, the event rate at 
low energies is artificially enhanced due to the SN rate which 
was assumed to be constant at high redshifts.

In the energy window from 19 to 35 MeV, the expected background 
event rate from the atmospheric $\antinue$ interacting with the 
protons ($\antinue p\rightarrow n e^+$) in the detector can be 
calculated. Using the atmospheric neutrino flux from Gaisser \etal 
\cite{gaisser}, the event rate for this background to the SRNs is 
only about $0.5/{\rm yr}$ (for 22.5 ktons of water). However, there 
is another source of background which is dominant. The atmospheric 
muon neutrinos interacting with the nucleons (both free and bound) 
in the fiducial volume produce muons. If these muons are produced 
with energies below Cerenkov radiation threshold (kinetic energy 
less than 53 MeV), then they will not be detected, but their 
decay-produced electrons and positrons will. Consequently, the muon 
decay signal will mimic the $\antinue p\rightarrow n e^+$ process 
in SuperK. The event rate from these muon decays was estimated 
 to be around unity for 0.58 kton-yr exposure of the 
Kamiokande II detector, forming the principal source of background 
after the various cuts had been implemented\cite{zhang}. Extrapolating to the 
fiducial volume of 22.5 ktons for SuperK, we expect that SuperK 
should see $\sim$39 events/yr as background to the SRN events. 
Although our predicted signal is much smaller than the sub-Cerenkov
muon background, it may still be detectable because the energy 
distributions of the signal and the background are distinctly different. 
In such a case, a conservative criterion for the detectability of the
signal is that it be greater than than the statistical fluctuations 
of the background.  However, even with three years of data {\em and} 
assuming that the SRN flux is close to our upper bound, the SRN signal 
is only just about equal to the statistical fluctuations in the 
sub-Cerenkov muon background. This situation will improve, though not 
dramatically, if SuperK can lower its threshold (to SRN) to 15 MeV.

Lastly we mention the SNO detector. Although much smaller than 
SuperK, the 1 kton SNO hopes to detect the SRN background by using 
the unique 2 neutron final state in $\antinue D \rightarrow nne^+$.  
Using the cross section of Kubodera and Nozawa~\cite{kubo}, the 
upper bound to the event rate above 10  MeV is a not-very-promising 
0.1/yr/kton. Again we show the differential event rate in Figure 
\ref{spectrum}.  Note, however, that unlike the SRN signal in 
Super-K, this rate can be influenced by the large $z$ SN rate, 
about which we know little.

\section{Discussion and Conclusions}\label{discuss}

\subsection{Previous works}

Supernova relic neutrinos have been the focus of many previous studies
\cite{totsuka,zhang,woosley,bis-kogan,krauss,tot-sato,dar,totani,malaney,hartmann}.  The fluxes predicted in these studies spread over some two 
orders of magnitude, primarily due to the uncertain determinations of the 
present number density of galaxies, the SN rate in our galaxy at present, 
and/or the SN redshift distribution.  More recently, Totani \etal\cite{totani} 
used the population synthesis method to model the evolution of star-forming 
galaxies and they obtained a prediction for the flux of SRN.  They found 
an event rate at SuperK (in the energy interval from 15 to 40 MeV) of 
$1.2~{\rm yr}^{-1}$ and the ``most optimistic'' prediction for their model 
was an event rate of 4.7/yr.  Malaney~\cite{malaney} used the Pei \& Fall 
results in order to parameterize the evolution of the cosmic gas density 
which he then uses to calculate the star formation rate and, from that, the
past supernova rate, finding a total SRN flux, integrated over all energies, 
of $2.0-5.4~{\rm{\mbox cm^{-2}sec^{-1}}}$ depending on somewhat arbitrary 
low redshift corrections to the supernova rate.  The work of Hartmann 
\& Woosley \cite{hartmann}, using a SN rate proportional to $(1+z)^4$ 
(motivated by \cite{pei,lilly})and normalized to the SN rate at present 
as derived from the $H\alpha$ observations of the local Universe is most 
similar to ours.  Their ``best" estimate of a relic neutrino flux is 
$\sim0.2~{\rm{\mbox cm^{-2}sec^{-1}}}$.  Although they do utilize Pei 
\& Fall beyond $z \sim 1$, they (and we and others) have noted that 
this contribution is subdominant. However, Hartmann \& Woosley \cite
{hartmann} do not discuss the backgrounds to detecting the SRN and 
although their estimated flux is smaller than our upper bound by about 
a factor of five, they conclude the SRN may somehow be detectable.
Although we agree with the Hartmann \& Woosley estimate of the SRN
flux in the sense that if we adopted their choices of parameters
rather than our ``conservative" choices we would predict the same
flux, we disagree that this small flux is detectable.

All these previous results are similar to, while less than 
the upper bound obtained in this paper. In fact, if we use our 
analysis of the SN rate along with the same IMF as used by Totani 
\etal for their spiral galaxies (which harbor most of the Type II 
supernovae~\cite{tammann,totani}), our SRN event rate in the 15 to 
40 MeV range (for comparison with Totani \etal) at SuperK falls to 
1/yr. The total integrated flux falls to $11~{\rm {\mbox cm^{-2}
sec^{-1}}}$ while the result for the flux in the 15 to 40 MeV energy 
window becomes $0.5~{\rm {\mbox cm^{-2}sec^{-1}}}$. These estimates agree 
well with those quoted from the previous works \cite{totani,malaney,hartmann}. 
In fact, the value of 1 event/yr obtained using the IMF from Totani \etal 
amounts to choosing the variables, $\mout$, $\aheat$ and $\aenu$ for an 
actual IMF rather than the extrema we have selected.  It is more likely 
that any realistic IMF (chosen to fit other observables) when combined 
with a SN rate that peaks at $z\sim 1$ (as implied by the metal enrichment 
history) will yield an {\it event rate that is  an order of magnitude 
smaller than the upper bound we quote}. Our upper-bound is robust because 
it is derived directly from the metal enrichment history which suggests 
that the \snii rate can peak no earlier than $z \sim 1$. 

\subsection{Choice of Cosmology}

Throughout, we have assumed that $\Omega_0=1$ and $q_0=0.5$. It is 
of some interest to ask how our SRN background predictions change 
if we change the background cosmology. Reducing the non-relativistic 
matter density from critical ($\Omega_0 < 1$), by allowing positive 
curvature and/or a cosmological constant $\Lambda$, would reduce the 
expansion rate at late times and thereby increase the SRN flux, for 
the {\it same} $H_0$. 
The event rate increases by about 40\%  in going from an $\Omega_0=1$ 
to an $\Omega_0=0.3,\,\Omega_\Lambda=0.7$ Universe. But the estimation 
of the luminosity density (which is used to derive the metal enrichment 
rates) itself requires the assumption of a background cosmology and 
it typically increases less rapidly with redshift for cosmologies 
with smaller $\Omega_0$~\cite{lilly}. These two effects tend to cancel 
out leaving the expected event rate nearly unchanged.  
Thus we do not expect our results to change substantially for a 
different background cosmology.

\subsection{Neutrino Oscillations}

The main goal of our work has been to obtain the most optimistic 
estimate of the SRN event rate at SuperK with the intent that if 
results from SuperK should exceed this upper bound, it could provide 
hints of new physics beyond the standard model.  Here, we consider 
neutrino oscillations as a mechanism for maximizing the SRN flux. 
Since $\bar{\nu}_{\rm x}$ (where x$=\mu\:{\rm or}\:\tau$) only 
experience neutral current interactions, they decouple deeper in 
the SN where the temperature is higher. As a result, they stream 
out of the SN with a higher temperature than the $\bar{\nu}_{\rm e}$. 
Because higher energy neutrinos are easier to detect, $\ebtoxb$ 
oscillations have the potential to increase the SRN event rate.  
The maximum effect for any scenario is attained when the mixing 
is maximal. We will assume a mass hierarchy wherein the electron 
neutrino is the lightest (however, for an inverted mass hierarchy 
and resonant conversion in the presence of magnetic fields, see
~\cite{totani2}). This implies that the MSW resonance condition 
is not satisfied for $\antinue\leftrightarrow\bar{\nu}_{\rm x}$, 
but vacuum oscillations can still occur. If all three flavors are 
maximally mixed, then the oscillation probabilities average out to 
1/3 for any reasonable choice of mass differences because of the 
large distances traversed by the the relic neutrinos (typically of 
order of $H_0^{-1}$; for the oscillation length to be comparable to 
this, $\Delta m^2 \sim 10^{-25}$ eV$^2$).  Such oscillations would 
make two-thirds of the original $\antinue$ flux hotter as they would 
be ``born'' (would have oscillated from $\bar{\nu}_{\rm x}$) with the 
same temperature as the $\bar{\nu}_{\rm x}$.  To quantify the discussion 
here, we take $\langle{\rm T}_{\bar{\nu}_{\rm x}}\rangle=2\langle{\rm T}
_{\antinue}\rangle$ (we might be exaggerating the spectral difference 
between $\bar{\nu}_{\rm x}$ and $\antinue$ considerably here~\cite{janka}) 
and assume that the same amount of energy is expelled in all three 
flavors. This leads to an upper bound to the SRN event rate at SuperK 
of 11/yr with an observable flux of $4~{\rm{\mbox cm^{-2}sec^{-1}}}$ 
in the 19 -- 35 MeV energy window.  The upper bound is larger by about 
a factor of 3 as a result of the increase in the number of neutrinos in 
the exponential tail of the neutrino distribution (where the observable 
energy window lies), due to the increase in temperature. As before, $\mout$ 
has been set equal to $1~\msun$. For the case where $\antinue$ is maximally 
mixed with only one of either $\bar{\nu}_\mu$ or $\bar{\nu}_\tau$, the 
upper bounds are 9/yr and  $3~{\rm {\mbox cm^{-2}sec^{-1}}}$ for the 
event rate and observable flux respectively. The upper bounds in the 
15 -- 35 MeV energy window are 14/yr for the three neutrino maximal 
mixing case, and 12/yr for the two neutrino maximal mixing case.  In 
general, a decrease in the threshold (below 19 MeV) {\em and} neutrino 
oscillations seem to be required to boost the SRN flux to sufficient 
levels.  Because the spectral shape of this oscillation enhanced SRN 
signal is sufficiently different from the sub-Cerenkov muon background, 
it may be detectable as a distortion of the expected muon background, 
{\em if} the SRN flux is in the vicinity of the upper bound we have 
quoted.  For SNO, the two neutrino maximal mixing case gives an event 
rate of 0.25/yr/kton while the three neutrino maximal mixing increases 
the rate to 0.29/yr/kton. 

A point to clarify here concerns the selection of the observable 
energy window given that the event rate in 19 -- 35 MeV window is 
now relatively large. Due to the larger signal, the relic neutrinos 
only become sub-dominant to the atmospheric neutrinos around 60 MeV
(for a relic neutrino flux close to the upper bound).  In fact, 
integrating out to a neutrino energy of order 60 MeV would increase 
the SRN event rate by about 50\%.  However the background from muon 
decay would increase by more than a factor of 3.  One possibility 
this opens up is to also use the energy window from about 55 to 70 
MeV since the muon-decay background is cut off at $m_\mu/2$ (as 
decay occurs for muons at rest).  For the same oscillation parameters 
as before, the upper bound to the event rate in the 55 to 70 MeV 
energy window is about 1/yr. However, the event rate due to the 
atmospheric neutrinos in the same energy window is of comparable 
magnitude and so, the situation is still not promising.

\subsection{Conclusions}

Using those observations most closely connected to the metal enrichment 
history of the Universe in order to relate the MER to the SNR, we have 
derived a robust upper bound to the supernova relic neutrino events at 
SuperK: 4 events in the energy window from 19 to 35 MeV for a 22.5 
kton-yr exposure.  We have argued that the SuperK signal is dominated 
by \snii from $z<1$ and so it is insensitive to the high redshift 
behavior of the metal enrichment rate. We use only the generic features 
of gravitational collapse SN models which have been substantiated by 
observations of SN 1987A to characterize the $\antinue$ spectrum 
emergent from \snii. In combination, these facts argue for the 
robustness of the upper bound to the SRN event rate obtained here. 
In addition, we have analyzed the backgrounds to the SRN events 
and conclude it is unlikely that SuperK will be able to detect 
these SRN neutrinos, unless the Type II supernova rate does not 
track the metal enrichment rate, or the observations of the star 
formation rate which lead to estimates of the metal enrichment rate 
at $z<1$ are in considerable error, and/or some physics beyond the 
standard model is at play.  We also find that the event rate at SNO 
will most likely be too small to be detected. The effect of flavor 
oscillations on the SRN flux has also been studied and the maximum 
possible increase in the event rate is less than a factor of 3.  
If the original flux is close to the upper bound quoted here and 
the mixing close to maximal, SuperK just might see the SRN flux 
as a distortion in its sub-Cerenkov muon background.
 
\subsection{Acknowledgments}

We acknowledge DOE support at The Ohio State University (DE-AC02-76ER01545) 
and at University of Chicago (DE-FG02-90ER40560).  We thank John Beacom for
useful discussions and for pointing out an error in an earlier version of 
this paper.

\begin{thebibliography}{5}
\bibitem{totsuka}Totsuka, Y. 1992, Rep. Prog. Phys. {\bf 55}, 377
\bibitem{sno} Boger, J. \etal (The SNO Collaboration) 1999, nucl-ex/9910016.
\bibitem{zhang}Zhang, W., \etal 1988, Phys. Rev. Lett. {\bf 61}, 385
\bibitem{sn1987a}Hirata, K., et. al. 1987, Phys. Rev. Lett. {\bf 58}, 1490
\bibitem{pei}Pei, Y.C. \& Fall, S.M. 1995, ApJ  {\bf 454}, 69
\bibitem{fall}Fall, S.M. `HST and the High Redshift Universe', (Proceedings
of the 37th Herstmonceaux Conference, held in Cambridge, UK, July 1996)
\bibitem{woosley}Woosley, S.E., Wilson, J.R. \& Mayle, R. 1986, 
ApJ {\bf 302}, 19
\bibitem{ww}Woosley, S.E. \& Weaver, T.A. 1995, ApJ Supp. {\bf 101}, 181
\bibitem{arnett}Arnett, W.D., Schramm, D.N. \& Truran, J.W. 1989, ApJ 
{\bf 339}, L25
\bibitem{smith}Pettini, M., Smith, L. J., King, D.L. 
\& Hunstead, R.W. 1997, ApJ {\bf 486}, 665 
\bibitem{lilly}Lilly, S.J., Le F$\acute{\rm e}$vre, O., Hammer, F. 
\& Crampton, D. 1996, ApJ {\bf 460}, L1
\bibitem{connolly}Connolly, A. J., Szalay, A. S., Dickinson, M., 
SubbaRao, M. U. \& Brunner, R. J. 1997, Astrophys. J. {\bf 486}, L11
\bibitem{gallego}Gallego, J., Zamorano, J., Arag$\acute{\rm o}$n-Salamanca, 
A. \& Rego, M. 1995, ApJ {\bf 455}, L1
\bibitem{madau}Madau, P., Ferguson, H.C., Dickinson, M.E., Giavalisco, M.,
Steidel, C.C. \& Fruchter, A. 1996, MNRAS {\bf 283}, 1388
\bibitem{boyle}Boyle, B. J. \& Terlevich, R. J. 1998, MNRAS {\bf 293}, L49
\bibitem{pascarelle}Pascarelle, S. M., Lanzetta, K. M. \& 
Fern$\acute{\rm a}$ndez-Soto, A. 1998, ApJ Lett. {\bf 508}, L1 
\bibitem{hughes} Hughes, D. H., \etal 1998, Nature {\bf 394}, 241
\bibitem{meurer}Meurer, G. R. \& Heckman, T. M. 1999, ApJ, to be published,
astro-ph/9903054. 
\bibitem{steidel}Steidel, C.  C., Adelberger, K. L., Giavalisco, M.,
Dickinson, M. \& Pettini, M. 1999, ApJ, to be published, astro-ph/9811399. 
\bibitem{lagage}Lapage, P.O. 1985, Nature {\bf 316}, 420 
\bibitem{gaisser}Gaisser, T.K., Stanev, T. \& Barr, G. 1988, 
Phys. Rev. D {\bf 38}, 85
\bibitem{vogel+beacom}Vogel, P. \& Beacom, J. F. 1999, 
Phys. Rev. D {\bf 60}, 053003
\bibitem{kubo}Kubodera, K. \& Nozawa, S. 1994, Int. J. Mod. Phys. E 
{\bf 3}, 101
\bibitem{bis-kogan}Bisnovatyi-Kogan, G.S. \& Seidov, Z.F. 1984, Sov. Astron., 
{\bf 26}, 132
\bibitem{krauss}Krauss, L.M., Glashow, S.L. \& Schramm, D.N. 1984, Nature 
{\bf 310},191
\bibitem{tot-sato}Totani, T. \& Sato, K. 1995, Astroparticle Phys. {\bf 3}, 
367
\bibitem{dar}Dar, A. 1985, Phys. Rev. Lett. {\bf 55}, 1422
\bibitem{totani}Totani, T., Sato, K. \& Yoshii, Y. 1996, ApJ {\bf 460}, 303
\bibitem{malaney}Malaney, R.A. 1997, Astroparticle Phys. {\bf 7}, 125 
\bibitem{hartmann}Hartmann, D. H. \& Woosley S. E. 1997, Astroparticle Phys.
{\bf 7}, 137
\bibitem{tammann}Tammann, G.A., L$\ddot{\rm o}$ffler, W. 
\& Schr$\ddot{\rm o}$der, A. 1994, ApJ Supp. {\bf 92}, 487
\bibitem{totani2}Totani, T. \& Sato, K. 1996, 
Int. J. Mod. Phys. D{\bf 5}, 519
\bibitem{janka}Janka, H.-T. 1995, Astropart. Phys. {\bf 3}, 377


\end{thebibliography}
\end{document}